\documentclass[11pt]{article}
\usepackage{blois,epsfig}
\usepackage{amsmath}
\usepackage{amsfonts}
\usepackage{amssymb}
\def\la{\lambda}
\def\al{\alpha}

\def\si{\sigma}

\def\be{\begin{equation}}
\def\ee{\end{equation}}
\def\bea{\begin{eqnarray}}
\def\eea{\end{eqnarray}}

\begin{document}
\vspace*{4cm}
\title{AXION BREMSSTRAHLUNG}
\author{D.V. GAL'TSOV$^*$,  E.Yu. MELKUMOVA$^*$ and R. KERNER$^\dag$}
\address{$^*$Department of Physics, Moscow State University,\\ 119899,
Moscow, Russia\\$^\dag$Universit\`e
Pierre et Marie Curie, 4, place Jussieu,\\
Paris, 75252, France}
 \maketitle
 \abstract { A new mechanism of cosmic
axion production is proposed: axion bremsstrahlung from collisions
of straight global strings. This effect is of the second order in
the axion coupling constant, but the resulting cosmological
estimate is likely to be of the same order as that corresponding
to radiation from oscillating string loops. This may lead to  a
further restriction on the axion window.}
%%%%%%%%%%%%%%%%%%%%%%%%%%%%%%%%%%%%%%%%%%%%%%%%%%%%%%%%%%%%%%%%%%%%%%
%%%%    Introduction                                              %%%%
%%%%%%%%%%%%%%%%%%%%%%%%%%%%%%%%%%%%%%%%%%%%%%%%%%%%%%%%%%%%%%%%%%%%%%
%

\section{Introduction}

The cosmic axion remains one of the viable candidates  for dark
matter, though the range of masses, which remains open accounting
for both collider and cosmological constraints, is rather narrow.
The model contains an unknown parameter, the vacuum expectation
value ${f}$, marking the energy scale of the U(1) symmetry
breaking~\cite{PQ77,W78,Wi78}. The mass of the axion, which is
acquired after the QCD phase transition, is inversely proportional
to $f$, whose upper bound is of  cosmological origin and follows
from the requirement that axions produced during the cosmological
evolution do not overclose the Universe. The lower bound on the
axion mass lies between several units and several tens of $\mu$eV
; this corresponds to the value of ${f}$ between $10^{12}$ and
$10^{11}$ GeV. For a recent survey of the present theoretical and
astrophysical status of the cosmic axion model
see~\cite{Sr02,Si02}, earlier reviews include~\cite{Re}.

The axion string network~\cite{HiKi95,BaSh97} is formed at the
temperature of the Peccei-Quinn phase transition ${f}$, and it is
usually assumed that the reheating temperature is higher than
${f}$, otherwise the network would be diluted by inflation.
Strings are primarily produced as long straight segments whose
length is of the order of the horizon size, and they initially
move with substantial friction~\cite{DS87,VI91,MaSh97} due to the
scattering  of cosmic plasma particles. At some temperature
$T_{*}<{f}$ scattering becomes negligible, and the string network
enters the scale-invariant regime~\cite{HS91,Na97,YKY} when
strings form closed loops and move almost freely with relativistic
velocities~\cite{AlTu,Sh87}. The standard estimates of the axion
radiation from global strings are based on the assumption that the
main contribution comes from the oscillating string
loops~\cite{DS89,DaQu90,YKY99,HaChSi01}.

Here we discuss another mechanism  for axion radiation:
bremsstrahlung, which has to be produced in collisions of long
strings. The existence of such an effect can be demonstrated as
follows. Consider two infinite straight strings inclined with
respect to each other and moving in parallel planes. Due to
interaction via the axion field, strings will be deformed around
the point of minimal separation between them. The motion of this
point is not associated with  the propagation of any signal, and
the corresponding velocity may be arbitrary, in particular,
superluminal. In the second order of string-axion interaction the
superluminally moving deformation must produce Cerenkov axion
radiation.  A similar mechanism was earlier suggested for
gravitational radiation of local strings,  but in that case the
explicit calculations~\cite{GaGrLe93} have led to  a zero result.
The vanishing of gravitational radiation, however, has a specific
origin related to the absence of gravitons in $2+1$ gravity
theory. Indeed, as  was explained in \cite{GaGrLe93}, two crossed
superluminal strings can be ``parallelized'' by suitable
coordinate transformations and world sheet reparameterizations, so
that the collision of strings is essentially equivalent to the
collision of point particles in $2+1$ dimensions. In the case of
the global strings similar considerations lead us to the problem
of  the electromagnetic radiation of point charges in $2+1$
dimensions, which is not forbidden by dimensionality. Thus one can
expect a non-vanishing axion bremsstrahlung from collisions of
global strings.
%%%%%%%%%%%%%%%%%%%%%%%%%%%%%
%

\section{Strings interacting via axion field. Axion radiation.}

Consider a pair of relativistic strings
$x^{\mu}=x_{n}^{\mu}(\sigma^{a}),\;\; \mu=0,1,2,3,\;\;
\sigma_{a}=(\tau, \sigma),\;\; a=0,1,$ where $n=1,2$ is the index
labelling them. The 4-dimensional space-time is assumed to be flat
and the signature is $+,---$ (and $(+,-)$ for the string
world-sheets). Strings interact via the axion field $B_{\mu\nu}$
as described by the action~\cite{BaSh97}
\begin{align}
S  & = -\sum_{n=1,2}\int\Big( \frac{\mu_{n}^{0}}{2} \sqrt{-\gamma}
\gamma ^{ab}\partial_{a} x_{n}^{\mu}\partial_{b}
x_{n}^{\nu}\eta_{\mu\nu} + 2\pi
f_{n} B_{\mu\nu} \epsilon^{ab} \partial_{a} x_{n}^{\mu}\partial_{b} x_{n}%
^{\nu}\Big)d^{2} \sigma_{n} + {\nonumber}\\
&   + \frac{1}{6 }\int\limits H_{ \mu\nu\lambda}
H^{\mu\nu\lambda}d^{4}x.
\end{align}
Here $\mu_{n}^{0}$ are the (bare) string tension parameters,
${f}_{n}$ are their axion couplings ($n$-labelling helps to
control the perturbation expansions), the Levy-Civita symbol is
chosen as $\epsilon^{01}=1,\; \gamma_{ab}$ is the metric on the
world-sheets. The totally antisymmetric
axion field strength is defined as $H_{ \mu\nu\lambda}=\partial_{\mu}%
B_{\nu\lambda}+\partial_{\nu}B_{\lambda\mu}+\partial_{\la}B_{\mu\nu}$
with the Lorentz gauge ${\partial}_{\la}B^{\nu\lambda}=0$. The
corresponding potential two-form is the sum
$B^{\mu\nu}=B^{\mu\nu}_{1}+B^{\mu\nu}_{2}$ of contributions
$B^{\mu\nu}_{n}$ due to each string:
\begin{equation}
\eta^{\alpha\beta}{\partial}_{\al}{\partial}_{\beta}B_{n}^{\mu\nu}=-4\pi
J_{n}^{\mu\nu}, \label{eeq}
\end{equation}
where $J_{n}^{\mu\nu}= \frac{f_{n}}{2} \int\limits
V_{n}^{\mu\nu}\, \delta ^{4}(x-x_{n}(\sigma_{n}))\,d^{2}
\sigma_{n}$ and $V_{n}^{\mu\nu} = \epsilon^{ab} \partial_{a}
x_{n}^{\mu}\partial_{b} x_{n}^{\nu}$.

In the gauge $\gamma_{ab}=\eta_{ab} $  the renormalized strings
equations of motion  read :
\begin{equation}
\mu_{n}\left( {\partial}_{\tau}^{2}-{\partial}_{\si}^{2}\right) x^{\mu}%
_{n}=2\pi f_{n} H^{\mu\nu\lambda}_{n^{\prime}}V_{\nu\lambda}^{n^{\prime}%
},\quad n\neq n^{\prime}, \label{steq}
\end{equation}
where the field $H^{\mu\nu\lambda}_{n^{\prime}}$ corresponds to
contribution of the $n^{\prime}-$th string (no sum over
$n^{\prime}$). The corresponding constraint equations are
$\dot{x}^{2} + x^{\prime 2} = 0, \quad\dot{x}^{\mu }x^{\prime
\nu}\eta_{\mu\nu} = 0 $ for each string, where dots and primes
denote the derivatives over $\tau$ and $\sigma$. The
\emph{self-action} terms in the equations of motion diverge both
near the strings and at large distances, so two regularization
parameters $\delta$ and $\Delta$ have to be
introduced~\cite{BaSh95} which are absorbed by the classical
renormalization of the string tension as
\begin{equation}
\mu=\mu_{0}+2\pi f^{2}\log(\Delta/\delta). \label{mure}
\end{equation}
The ultraviolet cutoff length $\delta$ is of the order of the
string thickness $\delta\sim1/f$, while the infrared cutoff
distance $\Delta$ usually is chosen as the correlation length of
the string network. Assuming that such a renormalization is
performed, we are left with the same equations of motion
(\ref{steq}), now with the physical tension parameters $\mu_{n}$,
in which only the \emph{mutual} interaction terms should be used
in the right hand side.

Our calculation follows the approach of~\cite{GaGrLe93} and it
consists in constructing solutions of the string equations of
motion and the axion field iteratively in terms of the
string-axion field interaction constant (equal to ${f}$). We would
like to mention also a similar perturbative approach to solve the
problem of gravitating point masses in terms of geodesic
deviations~\cite{KHC}. In the zero order approximation the strings
are moving
freely, so the lowest order axion fields due to both strings ${^{}_{1}B}%
_{n}^{\mu\nu} $ describe their mutual interaction and do not
contain the radiative part. Substituting them to (\ref{steq}), we
obtain the first order deformations of the world-sheets. These are
used to build the first order source terms ${_{1}J}_{n}^{\mu\nu}$
in the axion field equations (\ref{eeq}) which generate the second
order axion fields ${^{}_{2}B}_{n}^{\mu\nu}$ already containing
the radiative parts. The radiation power can be computed as the
reaction work produced by the half sum of the retarded and
advanced fields
upon the source and presented in the standard form`\cite{GMK}%
\begin{equation}
\mathcal{E} = \ \frac1{\pi}\int
k^{0}\epsilon(k^{0})|{_{1}J}_{\alpha\beta }(k)|^{2}
\delta(k^{2})d^{4}k, \label{final}
\end{equation}
where $\epsilon$ is the sign function.
%%%%%%%%%%%%%%%%%%%%%%%%%%%%%%
\newline
\indent The final formula for the axion bremsstrahlung from the
collision of two global strings can be obtained analytically in
the case of the
ultrarelativistic collision with the Lorentz factor $\gamma=(1-v^{2}%
)^{-1/2}\gg1$. The spectrum has an infrared divergence due to the
logarithmic dependence of the string interaction potential on
distance, so a cutoff length $\Delta$ has to be introduced. The
final formula for the radiated energy per unit length of the
target string reads:
\begin{equation}
\frac{d\mathcal{E}}{dl} =\frac{16\pi^{5} f^{6}\kappa}{3
\mu^{2}}\phi(y),\quad
y=\frac{d}{\gamma\kappa\Delta},\quad\kappa=\gamma\cos\alpha,
\end{equation}
here $\alpha$ is the strings inclination angle and
\[
\phi(y)=12\sqrt{\frac{y}{\pi}}\, {_{2}F_{2}}\left( \frac12,\frac
12;\frac32,\frac32;-y\right)  -3\ln\left( 4y\mathrm{e}^{C}\right)
+\frac72,
\]
where the generalized hypergeometric function is introduced and
$C$ is the Euler's constant. The leading term for small $y$ is the
logarithm, while the asymptotic value for large $y$ is constant
$\phi(\infty)=7/2$.

The velocity of the effective source (i.e. of the point of minimal
separation between the strings) tends to infinity when the string
become parallel. In this case, corresponding to $\kappa=\gamma$ in
the above formula, the Cerenkov cone opens up to $\theta=\pi/2$,
and the whole picture becomes essentially $2+1$ dimensional. This
gives an alternative description of the effect as bremsstrahlung
under collision of point electric charges in $2+1$
electrodynamics~\cite{GMK}.
%%%%%%%%%%%%%%%%%%%%%%%%%%%%%%%%%%%%%%%%%%%%%%%%%%%%%%%%%%%%%%%%%%%%%%%%%%%
%

\section{Cosmological estimates}

Our mechanism presumably works in the temperature interval
$(T_{*},\,T_{1})$ between the onset of the scaling regime till the
QCD phase transition, though an additional contribution form the
damped epoch $T_{0}<T<T_{*}$ can also be expected. The scaling
density of strings is determined from numerical experiments, it
can be presented as $\rho_{s}=\zeta\mu/t^{2} $ with $\zeta$
varying from 1 to 13 in different simulations. The corresponding
moments of time are commonly estimated as~\cite{BaSh97}
\begin{equation}
t_{*} (\mathrm{sec}) \sim10^{-20}\left(
\frac{f}{10^{12}\mathrm{GeV}}\right)
^{-4},\quad t_{1} (\mathrm{sec})\sim2\cdot10^{-7}\left( \frac{f}%
{10^{12}\mathrm{GeV}}\right) ^{1/3}. \label{t*}
\end{equation}
Consider the scattering of an ensemble of randomly oriented
straight strings on a selected target string in the rest frame of
the latter. Since the dependence of the string bremsstrahlung on
the inclination angle $\alpha$ is smooth, we can use for a rough
estimate the result for parallel strings ($\alpha=0$) introducing
an effective fraction $\nu$ of ``almost'' parallel strings
(roughly 1/3). Then, if there are $N$ strings in the normalization
cube $V=L^{3}$, we have to integrate the radiation energy released
in the collision with the impact parameter $d=x$ over the plane
perpendicular to the target string using the measure
$N/L^{2}\cdot2\pi x dx$. Actually we need the radiation power per
unit time, so, for an estimate, we have to divide the integrand on
the impact parameter. This quantity should be multiplied by the
total number of strings $N$ to get the radiation energy released
per unit time within the normalization volume. Therefore, for the
axion energy density generated per unit time we obtain:
\begin{equation}
\frac{d \varepsilon_{a}}{dt}=\int_{0}^{L} \mathcal{E}\,\nu\,\frac{N}{L^{2}%
}\,\frac{N}{V}\,2\pi dx.
\end{equation}
Substituting here the Eq. (\ref{final}), we integrate over the
impact parameter, taking into account that the string number
density is related to the energy density via ~ $N/V=\rho_{s}/\mu L
$. Then introducing the integral~$\xi(w)=\int_{0}^{w} \phi(y) dy $
and keeping only the second (leading) term in the expression
(\ref{mure}) for the string tension $\mu$, we obtain
\begin{equation}
\frac{d\varepsilon_{a}}{dt}\sim K  f^{2}\gamma^{3}%
\xi(w),\quad w=\frac{L}{\gamma^{2}\Delta},
\end{equation}
where $K=\frac83\pi^{4}\nu\zeta^{2}$. Since in the cosmological
context $L\sim\Delta\sim t$, one can take $w=\gamma^{-2}$ for an
estimate. The ``realistic'' value of $\gamma$ is of the order of
unity, while our formulas were obtained in the $\gamma\gg1$
approximation. But an independent calculation shows that for
$\gamma\sim1$ the contribution from the first term is of the same
order, so we can hope to get reasonable order of magnitude
estimate using the above formulas for $\gamma$ not very large. The
full $\gamma$-dependence is given by the function
$\gamma^{3}\xi(\gamma^{-2})$, which for $\gamma=\sqrt{2}$ is ~$\gamma^{3}%
\xi(\gamma^{-2})\approx 8.137,$~while for
$\gamma=5$,\,$\gamma^{3}\xi (\gamma^{-2})\approx 55.83$. Another
source of uncertainty is the numerical value of the parameter
$\zeta$, which was obtained in different simulations within the
region between 1 and 13.

Integrating the radiation power over time from $t_{*}$ to $t_{1}$
as given by (\ref{t*}) one finds the energy density of axions at
the moment of QCD phase transition $t_{1}\gg t_{*}$:
\begin{equation}
\varepsilon_{a} \sim\frac{K f^{2}\gamma^{3}\xi(w)}{2t_{*}^{2}\ln^{2}%
(t_{*}/t_{0})}.
\end{equation}
Finally, dividing this by
the critical energy density at $t=t_{1}$ ~$\varepsilon_{\mathrm{cr}}%
=\frac{3m_{\mathrm{Pl}}^{2}}{32\pi t_{1}},\quad
m_{\mathrm{Pl}}=1.22\cdot 10^{19}\mathrm{GeV}$, and using the
relation $t_{*}/t_{0}=(m_{\mathrm{Pl}}/f)^2$, we obtain for
$\gamma=\sqrt{2}$ the following estimate for the relative
contribution of the bremsstrahlung axions
\begin{equation}
\Omega^{\mathrm{br}}_{a}\sim 0.5 \ 10^{16}\left(
\frac{\zeta}{13}\right) ^{2}\left(
\frac{f}{10^{12}\mathrm{GeV}}\right) ^{32/3}.
\end{equation}
This rough estimate shows that our new mechanism is very sensitive
to the value of $f$ and gives an upper bound on  the axion window
of the order of several $10^{10} \mathrm{GeV}$. We postpone more
accurate cosmological estimates for a separate publication. In
particular, more careful analysis is needed to accommodate the
analysis to the actual astrophysical value of the string Lorentz
factor.
%%%% %%%%%%%%%%%%%%%%%%%%
%%%%%%%%%%%%%%%%%%%%%%%%%%%%%%%%%
%     REFERENCES
%%%%%%%%%%%%%%%
%

\end{document}